\documentclass[english,prb,twocolumn,notitlepage,superscriptaddress,showpacs]{revtex4-1}

\usepackage{graphicx}
\usepackage{color}
\usepackage{amssymb}
\usepackage{amsmath}
\usepackage{setspace}
\usepackage{babel}
\usepackage[utf8x]{inputenc}

\begin{document}

\title{Coherent Tunneling by Adiabatic Passage of an exchange-only spin qubit in a double quantum dot chain}
\author{E. Ferraro}
\email{e.ferraro@inrim.it}
\affiliation{Laboratorio MDM, IMM-CNR, Via Olivetti 2, I-20864 Agrate Brianza, Italy}
\altaffiliation[Present address: ]{Istituto Nazionale di Ricerca Metrologica, Strada delle Cacce 91, 10135 Torino, Italy}
\author{M. De Michielis}
\affiliation{Laboratorio MDM, IMM-CNR, Via Olivetti 2, I-20864 Agrate Brianza, Italy}
\author{M. Fanciulli}
\affiliation{Laboratorio MDM, IMM-CNR, Via Olivetti 2, I-20864 Agrate Brianza, Italy}
\affiliation{Dipartimento di Scienza dei Materiali, University of Milano Bicocca, Via R. Cozzi, 53, 20126 Milano, Italy}
\author{E. Prati}
\affiliation{Laboratorio MDM, IMM-CNR, Via Olivetti 2, I-20864 Agrate Brianza, Italy}
\affiliation{Istituto di Fotonica e Nanotecnologia, Consiglio Nazionale delle Ricerche, Piazza Leonardo da Vinci 32, I-20133 Milano, Italy}

\begin{abstract}
A scheme based on Coherent Tunneling by Adiabatic Passage (CTAP) of exchange-only spin qubit quantum states in a linearly arranged double quantum dot chain is demonstrated. Logical states for the qubit are defined by adopting the spin state of three electrons confined in a double quantum dot. The possibility to obtain gate operations entirely with electrical manipulations makes this qubit a valuable architecture in the field of quantum computing for the implementation of quantum algorithms. The effect of the external control parameters as well as the effect of the dephasing on the coherent tunneling in the chain is studied. During adiabatic transport, within a constant energy degenerate eigenspace, the states in the double quantum dots internal to the chain are not populated, while transient populations of the mixed states in the external ones are predicted.
\end{abstract}

\maketitle
A Coherent Tunneling by Adiabatic Passage (CTAP) scheme of exchange-only spin qubit quantum states in a linearly arranged double quantum dot chain is proposed to transfer the information in a quantum circuit by adopting Gaussian pulses.

Communication among distant qubits is one of the most challenging requirements towards the implementation of quantum algorithms in solid state system. CTAP in chains of quantum dots where logical states are encoded by one electron has been theoretically predicted in Refs.\cite{huneke,greentree,triple3,greentree2} and later experimentally demonstrated in GaAs \cite{vandersypen}. The growing interest in quantum computation pushes towards the realization of practical devices that interconnect remote sites composing the quantum circuit to transfer information. Here we extend the formalism of CTAP to the exchange only qubits introduced in Ref.\cite{shi} and further developed in Refs.\cite{koh,ncom,koh2,wis}. There, logical states are defined by adopting combined spin states of three electrons confined electrostatically in double quantum dots. In particular the logical states are expressed by $|0\rangle\equiv|S\rangle|\downarrow\rangle$ and $|1\rangle\equiv\sqrt{\frac{1}{3}}|T_0\rangle|\downarrow\rangle-\sqrt{\frac{2}{3}}|T_-\rangle|\uparrow\rangle$ where $|S\rangle$, $|T_0\rangle$ and $|T_-\rangle$ are respectively the singlet and triplet states of the spin of a pair of electrons embedded in one dot and $|\uparrow\rangle$ and $|\downarrow\rangle$ denote the spin-up and spin-down of the electron in the other dot. The effective Hamiltonian model for a single qubit was derived in Ref.\cite{ferraro}, the corresponding one for the two qubits case was recently developed in Ref.\cite{ferraro2}, a universal set of quantum gates is presented in Ref.\cite{mdm1} and an implementation based on Si-MOS quantum dots compatible with the CMOS industrial technological standards is designed in Ref.\cite{RottaQIC2014}. The CTAP scheme in our case consists in the tunneling of the three electrons localized initially in the first double quantum dot \cite{pierre,shi} at the head of the chain to the end by using all electrical manipulations. The adiabatic passage takes place when a quantum system prepared in an arbitrary superposition of eigenstates of the Hamiltonian changes slowly during the course of time, that is the transitions between eigenspaces are negligible \cite{greentree2, rigolin}. In particular it is demonstrated that for the chain under study the adiabatic tunneling takes place within the degenerate eigenspace with constant energy. Moreover population transfer between two distant double quantum dots is simulated in a double quantum dot chain without occupation in the internal quantum dots. A workable Hilbert space is provided by both GaAs \cite{see,mac} and Si \cite{prati} quantum dots, where the latter further requires sufficient energy separation of valley orbitals \cite{mdm2}.


\section{Transport in a double quantum dot chain}
The Hamiltonian model of the chain where CTAP is imposed is introduced. Next the dynamical problem is solved through a master equation approach.

\begin{figure}[h]
\begin{center}
\includegraphics[width=0.5\textwidth]{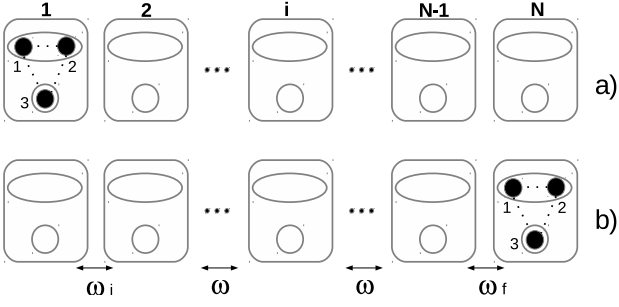}
\end{center}
\caption{Schematic of the tunneling in the quantum channel with $N$ double quantum dots. The black dots in (a) and (b) denote the electrons before and after the tunneling. The quantities indicated by $\omega_i$, $\omega$ and $\omega_f$ are the pulses that have to be furnished by external gates in order to make the tunneling possible.}\label{Fig1}
\end{figure}

A schematic of the quantum channel made of an odd number $N$ of identical double quantum dots is reported in Figure \ref{Fig1}, where the electrons denoted by the black dots before (a) and after (b) the tunneling have been depicted.

The Hamiltonian model that describes the system under investigation is expressed by the sum of the following three terms in units of $\hbar$:
\begin{equation}\label{ham}
H=\sum_{i=1}^NH_i+W(t),
\end{equation}
where $N$ is the number of the double quantum dot composing the channel. The first terms
\begin{align}
H_i=&\sum_{k=1}^3\sum_{\sigma}\varepsilon_k^ic_{k\sigma}^{i\dagger}c_{k\sigma}^{i}+\sum_{k=1}^3\sum_{\sigma}U_k^in_{k\uparrow}^in_{k\downarrow}^i+\nonumber\\
&+U_{12}^i(n_{1\uparrow}^i+n_{1\downarrow}^i)(n_{2\uparrow}^i+n_{2\downarrow}^i)
\end{align}
are the free Hamiltonians of each double quantum dot that include the single electron energy level of each quantum dot and the Coulomb interactions \cite{shi}. The terms that express the application of gate voltages for the coherent tunneling are given by:
\begin{align}
&W(t)=W^{(1)}(t)+W^{(2)}(t)\nonumber\\
&W^{(1)}(t)=-\sum_{i=1}^{N-1}\omega_{i,i+1}(t)\sum_{\sigma}c_{3\sigma}^{i\dagger}c_{3\sigma}^{i+1}+h.c.\nonumber\\
&W^{(2)}(t)=-\sum_{i=1}^{N-1}\omega_{i,i+1}(t)\sum_{\sigma\neq\sigma'}c_{1\sigma}^{i\dagger}c_{1\sigma}^{i+1}c_{2\sigma'}^{i\dagger}c_{2\sigma'}^{i+1}+h.c.
\end{align}
where the first order term refers to the tunneling of the electron embedded in one quantum dot \cite{huneke} and the second order term refers instead to the simultaneous passage of the pair of the electrons contained in the second quantum dot. Other first order transitions involving the electrons in the doubly occupied quantum dot are forbidden by selection rule on the energy.

\begin{figure}[h]
\begin{center}
\includegraphics[width=0.5\textwidth]{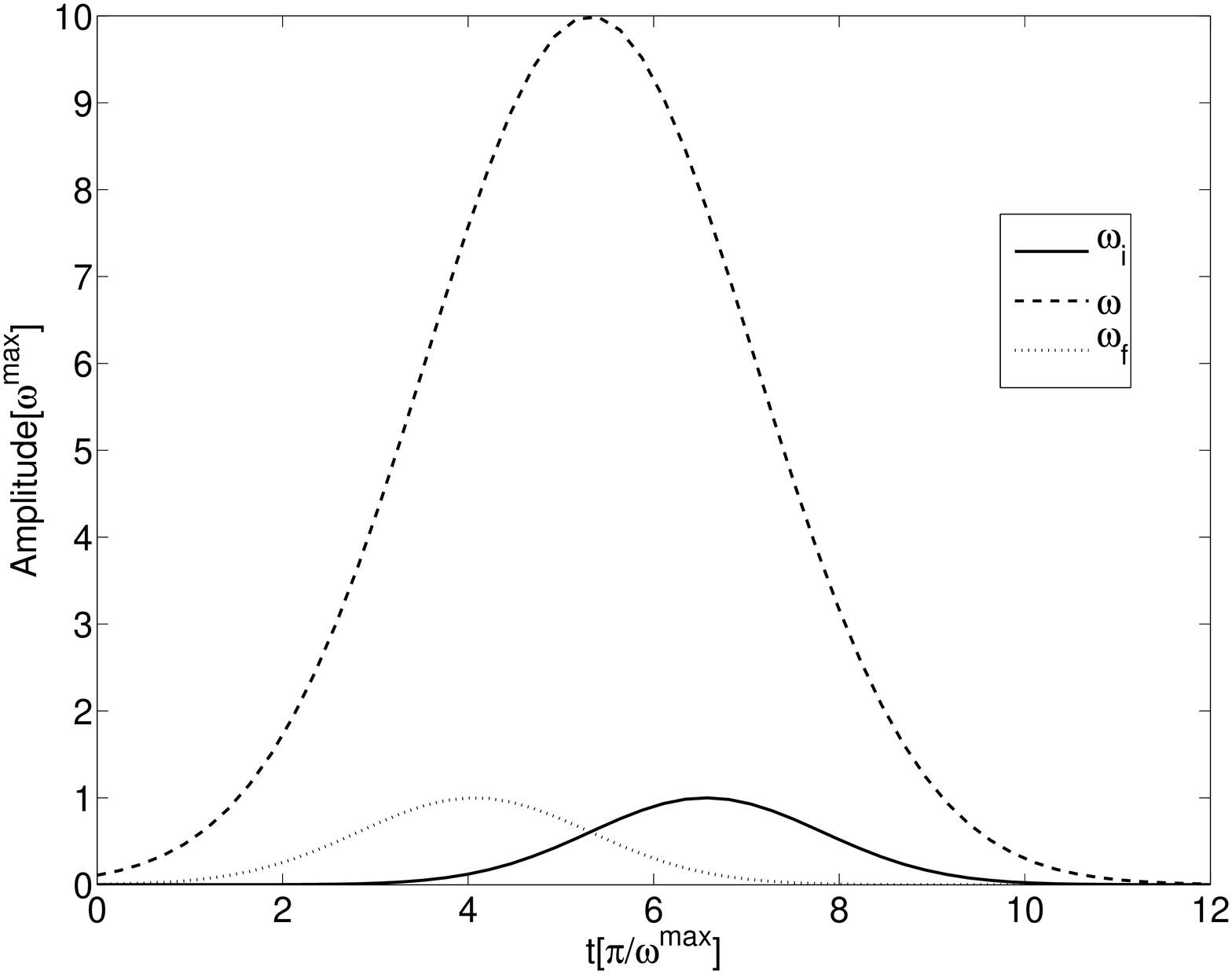}
\end{center}
\caption{Gaussian pulses as a function of $t$, for $t_{max}=10\pi/\omega^{max}$, $\sigma=t_{max}/8$ and $\omega^{max}_S=10\omega^{max}$}\label{Fig2}
\end{figure}

The series of pulses $\omega_{i,i+1}(t)$ considered in the framework of adiabatic tunneling are of Gaussian shape \cite{greentree}. In particular the intermediate pulses are equal among them but distinct from the external ones that are applied in a counter-intuitive sequence \cite{greentree}, that is from the tail to the head of the chain in such a way that $\omega_f$ is applied before $\omega_i$. It has a significant advantages in enabling high fidelity transfer. The pulses are explicitly expressed by
\begin{align}\label{Eq:sequence}
&\omega_i\equiv\omega_{1,2}(t)=\omega^{max}exp\left\{-\frac{[t-(t_{max}/2+\sigma)]^2}{2\sigma^2}\right\}\nonumber\\
&\omega\equiv\omega_{i,i+1}(t)=\omega^{max}_Sexp\left\{-\frac{(t-t_{max}/2)^2}{4\sigma^2}\right\} \forall\;2\le i<N-1\nonumber\\
&\omega_f\equiv\omega_{N-1,N}(t)=\omega^{max}exp\left\{-\frac{[t-(t_{max}/2-\sigma)]^2}{2\sigma^2}\right\},
\end{align}
where $t_{max}/2\pm\sigma$ $(t_{max}/2)$ is the peak time and $\sigma$ $(\sqrt{2}\sigma)$ is the standard deviation of the external (intermediate) double quantum dots composing the chain. $\omega^{max}$ ($\omega_S^{max}$) is the maximum tunnelling rate for the transitions between states of the external (intermediate) double quantum dots. Figure \ref{Fig2} shows the shape of the Gaussian pulses adopted in the following, where the intermediate pulse $\omega^{max}_S$ has an amplitude 10 times greater than the amplitude $\omega^{max}$ of the external pulses to achieve high fidelity transfer as reported in Ref.\cite{greentree} and where the values for $t_{max}$ and $\sigma$ are fixed.

We point out that in our case the last double quantum dot of the chain is tuned identically to the first and, during the passage, the chain forms a single quantum system. During the time evolution of such single system, there is no heat exchange and energy remains constant. The final state cannot be made of two electrons in two different dots, as the Coulomb repulsion energy contributes to the energy in the initial state. For this reason the two electrons are coherently transported across the chain together and processes leading to alternative final states are forbidden by energy conservation.

Let's solve the dynamical problem by starting from Hamiltonian (\ref{ham}). A master equation approach for the density matrix $\rho(t)$ describing the total system composed by $N$ double quantum dot has been adopted. The master equation \cite{greentree}
\begin{equation}\label{me}
\dot{\rho}(t)=-\frac{i}{\hbar}[W(t),\rho(t)]-\Gamma[\rho(t)-diag(\rho(t))]
\end{equation}
where $\Gamma$ is the pure dephasing rate, assumed to act equally on all coherences, has been solved numerically.

\begin{figure}[h]
\begin{center}
\includegraphics[width=0.5\textwidth]{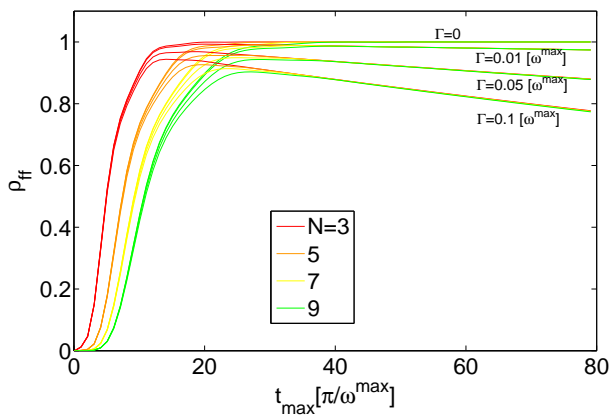}
\end{center}
\caption{Probability $\rho_{ff}$ of finding the three electrons at the tail of the chain in the double quantum dot $N$ as a function of $t_{max}$ for $\sigma=t_{max}/8$ and $\omega^{max}=10$. Each family of curves represents a different value of the dephasing parameter $\Gamma$ for a different number of double quantum dots composing the chain.}\label{nhq}
\end{figure}

In Figure \ref{nhq} the solutions of the master equation (\ref{me}) are shown for different values of the parameter $\Gamma$ and for different values of the number $N$ of double quantum dots composing the chain. Starting from the initial condition in which all the three electrons are initialized in the first double quantum dot, we plot the element $\rho_{ff}$ of the density matrix representing the probability of finding the three electrons at the tail of the chain in the double quantum dot $N$. We have chosen to report the results as a function of the Gaussian parameter $t_{max}$, by fixing the standard deviation to $\sigma=t_{max}/8$ and $\omega^{max}=10$.

In the ideal case, represented by $\Gamma=0$, perfect transfer is obtained for $t_{max}$ greater than a value that increases with $N$.
For $N=9$ a total pulse time $t_{max}$ greater than $40[\pi/\omega^{max}]$ guarantees the complete transfer.

When dephasing is added ($\Gamma>0$) a perfect transfer can not be achieved. $\rho_{ff}$ has a maximum ($\rho_{ff}^{max}$) at a value of $t_{max}$ which varies with $N$. As expected, $\rho_{ff}^{max}$ decreases when $\Gamma$ is increased.
Setting $\omega^{max}= 10 \mu eV$, in the case of $\Gamma = 0.05 [\omega^{max}]$ = 0.76 GHz (close to the rate adopted in Ref. \cite{koh2}) and by considering a chain of $N=9$ double quantum dots, the optimum value for $t_{max}$ is $t_{max}$=$29.1 [\pi/\omega^{max}]$= 6.01 ns with $\rho_{ff}^{max}$= 0.94. The total transfer time is slightly longer than $t_{max}$ due to duration of the vanishing Gaussian tail of the broader pulses applied to the intermediate double quantum dots.
Another source of infidelity in the CTAP is the presence of miscalibration in the control signals $\omega_{i}(t)$, $\omega(t)$, $\omega_{f}(t)$ with respect the ideal series of Gaussian pulses. In Appendix \ref{App:Miscalibration}, the effects on the interaction Hamiltonian $W(t)$ of an unwanted deviation in the amplitude or in the peak time of the control signals from the ideal Gaussian pulses are investigated.

The CTAP is not the unique method which can be exploited to perform the transmission of quantum states between two remote qubits connected via a double quantum dot chain. A different approach is based on successive SWAP operations where the SWAP gate exchanges quantum states of two adjacent qubits, moving the quantum state of the head qubit towards the tail one in successive steps. The comparison between transfer performances of the two approaches is presented in Appendix \ref{App:CTAPvsSWAP}.

\section{CTAP in a chain of three double quantum dots}
A quantum channel made by a chain of three double quantum dots is here examined in more detail. In this quantum channel the matrix form of the interaction part of the Hamiltonian model is here reported:
\begin{equation}\label{h3}
\hspace{-0.3cm}W=\left(\begin{array}{ccccccccc}
0 & -\omega_i & 0 & -\omega_i & 0 & 0 & 0 & 0 & 0\\
-\omega_i & 0 & -\omega_f & 0 & -\omega_i & 0 & 0 & 0 & 0\\
0 & -\omega_f & 0 & 0 & 0 & -\omega_i & 0 & 0 & 0\\
-\omega_i & 0 & 0 & 0 & -\omega_i & 0 & -\omega_f & 0 & 0\\
0 & -\omega_i & 0 & -\omega_i & 0 & -\omega_f & 0 & -\omega_f & 0\\
0 & 0 & -\omega_i & 0 & -\omega_f & 0 & 0 & 0 & -\omega_f\\
0 & 0 & 0 & -\omega_f & 0 & 0 & 0 & -\omega_i & 0\\
0 & 0 & 0 & 0 & -\omega_f & 0 & -\omega_i & 0 & -\omega_f\\
0 & 0 & 0 & 0 & 0 & -\omega_f & 0 & -\omega_f & 0\\
\end{array}\right)
\end{equation}
with state ordering $\{|P_1S_1\rangle,|P_1S_2\rangle,|P_1S_3\rangle,|P_2S_1\rangle,|P_2S_2\rangle,\\|P_2S_3\rangle,
|P_3S_1\rangle,|P_3S_2\rangle,|P_3S_3\rangle\}$. In the basis chosen $P$ and $S$ stand respectively for the state of the pair of electrons in the same quantum dot and the state of the single electron in the other. The indices $1,2,3$ refer to the double quantum dot composing the quantum channel.

\begin{figure}[h]
\begin{center}
\includegraphics[width=0.5\textwidth]{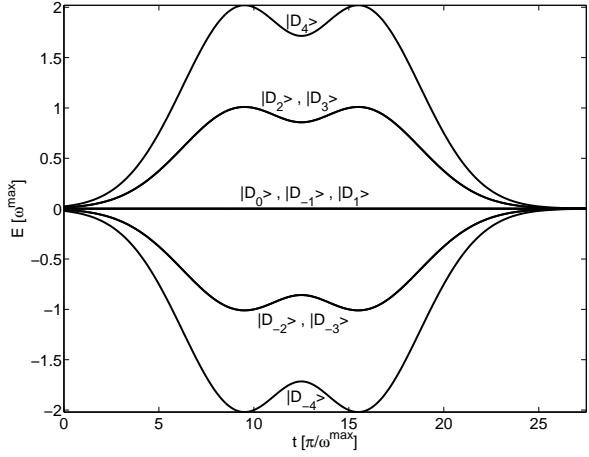}
\end{center}
\caption{Eigenvalues of three double quantum dots system as a function of time. The state at zero energy is triple degenerate.}\label{Fig:eigenval-t}
\end{figure}

The eigenvalues of the triple quantum dot system with pulse sequence defined by Eq.(\ref{Eq:sequence}) as a function of $t$ with $\omega^{max}=1$, $t_{max}=25\pi/\omega^{max}$ and $\sigma=t_{max}/8$ are reported in Figure \ref{Fig:eigenval-t}. Note that there are three degenerate states at zero energy during the entire pulse sequence, by granting the adiabatic passage to take place. When the eigenvalues are degenerate, adiabaticity ensures that the system stays within the eigenspaces of the initial eigenvalue. Contrarily from the non-degenerate case, the system remains in the same eigenstate when the Hamiltonian varies.

The three degenerate eigenstates of the Hamiltonian (\ref{h3}) in correspondence to zero energy are:
\begin{align}\label{eigen}
|D_0\rangle=&\frac{1}{\sqrt{2\omega_i^4+\omega_f^4}}\left[-(\omega_f^2-\omega_i^2)|P_1S_1\rangle+\omega_i\omega_f(|P_1S_3\rangle+\right.\nonumber\\
&\left.+|P_3S_1\rangle)-\omega_i^2|P_2S_2\rangle\right]\nonumber\\
|D_{-1}\rangle=&\frac{1}{\sqrt{2\left(\omega_i^2+\omega_f^2\right)}}\left[\omega_i(|P_1S_2\rangle-|P_2S_1\rangle)+\right.\nonumber\\&\left.-\omega_f(|P_2S_3\rangle-|P_3S_2\rangle)\right]\nonumber\\
|D_1\rangle=&\frac{1}{\sqrt{3}}(|P_1S_1\rangle-|P_2S_2\rangle+|P_3S_3\rangle).
\end{align}
A complete discussion of adiabatic evolution in degenerate subspaces can be found in \cite{rigolin}.

Let's demonstrate that a combination of the eigenstates (\ref{eigen}) guarantees the adiabatic tunneling. Any superposition of them belongs to the same zero energy eigenspace, in particular the following superposition is considered
\begin{equation}
|D\rangle\propto-|D_0\rangle+\frac{\omega_i^2}{\sqrt{2\omega_i^4+\omega_f^4}}\sqrt{3}|D_1\rangle
\end{equation}
that after the normalization has the following form
\begin{align}\label{do}
|D\rangle=&\frac{1}{\omega_i^2+\omega_f^2}\left[\omega_f^2|P_1S_1\rangle-\omega_i\omega_f(|P_1S_3\rangle+|P_3S_1\rangle)+\right.\nonumber\\
&\left.+\omega_i^2|P_3S_3\rangle\right].
\end{align}
The state (\ref{do}) is analogous to the states derived in Refs.\cite{greentree2, petro, jong, shore}.

\begin{figure}[h]
\begin{center}
\includegraphics[width=0.5\textwidth]{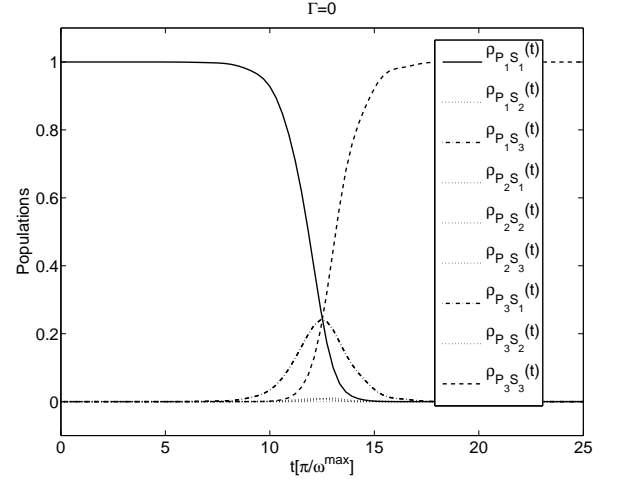}
\end{center}
\caption{Populations of the density matrix $\rho$ as a function of time for $N=3$, $\omega^{max}=1$, $t_{max}=25\pi/\omega^{max}$, $\sigma=t_{max}/8$ and $\Gamma=0$}\label{3h2d}
\end{figure}

In the adiabatic limit the passage is from the state $|P_1S_1\rangle$ when $t=0$ and $\omega_i\ll\omega_f$ to $|P_3S_3\rangle$ when $t \rightarrow+\infty$ and $\omega_i\gg\omega_f$ by maintaining the system in (\ref{do}), without population leakage into the other eigenstates and transient population in states $|P_1S_3\rangle$ and $|P_3S_1\rangle$. Such time evolution is shown in Figure \ref{3h2d} where the populations of the density matrix as a function of time starting from the initial condition $\rho_{P_1S_1}(0)=1$ is reported.

At the end of the process as expected $\rho_{P_3S_3}(0)=1$ while all the others are zero, meaning that an ideal transfer takes place. This results is true irrespective of the number of double quantum dot composing the chain, the central double quantum dot is always not occupied through the transit.

\begin{figure}[h]
\begin{center}
\includegraphics[width=0.5\textwidth]{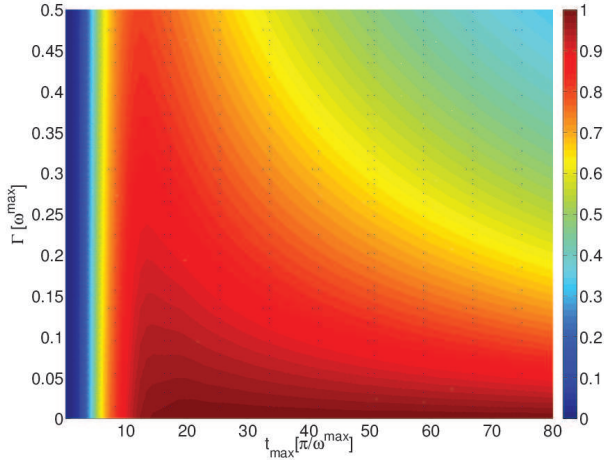}
\end{center}
\caption{Probability $\rho_{P_3S_3}$ of finding the three electrons at the tail of the chain for $N=3$ as a function of $t_{max}$ and $\Gamma$ for $\sigma=t_{max}/8$ and $\omega^{max}=10$.}\label{3h3d}
\end{figure}

The adiabatic tunneling with dephasing is analyzed by solving the master equation (\ref{me}) numerically. Figure \ref{3h3d} recaps in a contour plot the populations $\rho_{P_3S_3}$ when the dephasing $\Gamma$ and the peak time $t_{max}$ are varied simultaneously.
Within the region with $10[\pi/\omega^{max}]\le t_{max}\le40[\pi/\omega^{max}]$ and $\Gamma$ lesser than $0.1[\omega^{max}]$ a coherent tunneling in a three double quantum dots channel with $\rho_{P_3S_3}>0.9$ can be achieved.

\begin{figure}[h]
\begin{center}
\includegraphics[width=0.5\textwidth]{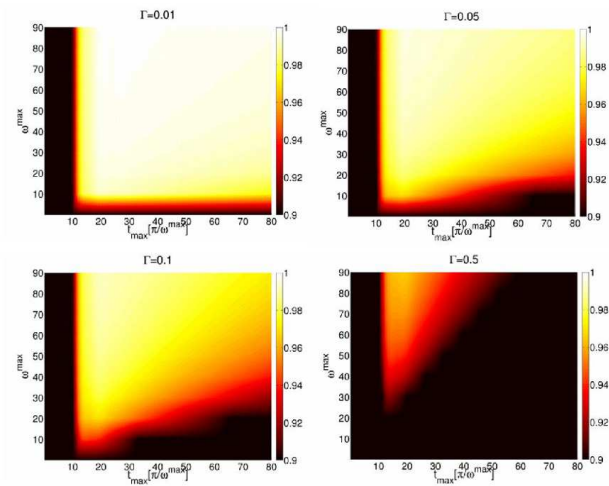}
\end{center}
\caption{Probability $\rho_{P_3S_3}$ of finding the three electrons at the tail of the chain for $N=3$ as a function of $t_{max}$ and $\omega^{max}$ for $\sigma=t_{max}/8$. Each plot illustrates a different value of the dephasing $\Gamma$.}\label{3hw}
\end{figure}

Moreover the dependence of the tunneling rates $\omega^{max}$ on the coherent tunneling is calculated. It is governed by the transitions between the states of the double quantum dots composing the chain which in turn are controlled through external gates. In Figure \ref{3hw} a collection of 3D plots is reported in which the results obtained by varying $\omega^{max}$ are shown for different values of the dephasing $\Gamma$.

Figure \ref{3hw} shows the condition that assures a good transfer, which is given mostly by a large value of $\omega^{max}$. It implies that also for a large dephasing, for example $\Gamma=0.1$, one can obtain an optimal transfer by choosing an $\omega^{max}$ ten times greater than in the $\Gamma=0.01$ case.

\section{Conclusions}
A CTAP scheme of exchange-only spin qubit quantum states in a double quantum dot chain is obtained. The scheme allows the communication between remote sites composing a quantum circuit based on three electrons spin states in double quantum dots. Unlike to single electron or singlet-triplet qubit, double quantum dot exchange-only spin qubit needs the transfer of three electrons that are used to encode the logical states. The passage from the initial state to the final one occurs in the adiabatic limit by maintaining the system in the zero energy degenerate eigenspace. Population leakage into the other eigenstates is absent and transient populations are observed in the three double quantum dot chain. Moreover the role played by the tunneling rates and by the dephasing on the coherent tunneling is analyzed. An increase in the dephasing parameter corresponds to a worsening in the coherent tunneling that however may be improved by choosing a tunneling rate greater than in the no-dephasing case.

\acknowledgements
This work is partially supported by the project QuDec, Italian Ministry of Defence.

\appendix
\section{Miscalibration analysis}
\label{App:Miscalibration}
\begin{figure}[h]
\begin{center}
\includegraphics[width=0.5\textwidth]{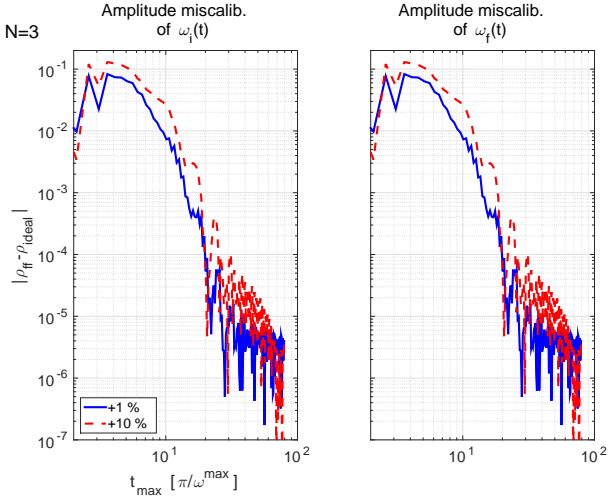}
\end{center}
\caption{Absolute value of the difference between $\rho_{ff}$ and $\rho_{ideal}$ as a function of $t_{max}$ for two different variation (+ 1\% and +10\%) on the amplitude of $\omega_{i}(t)$ and $\omega_{f}(t)$ for $N$=3.}\label{N3amp}
\end{figure}

\begin{figure}[h]
\begin{center}
\includegraphics[width=0.5\textwidth]{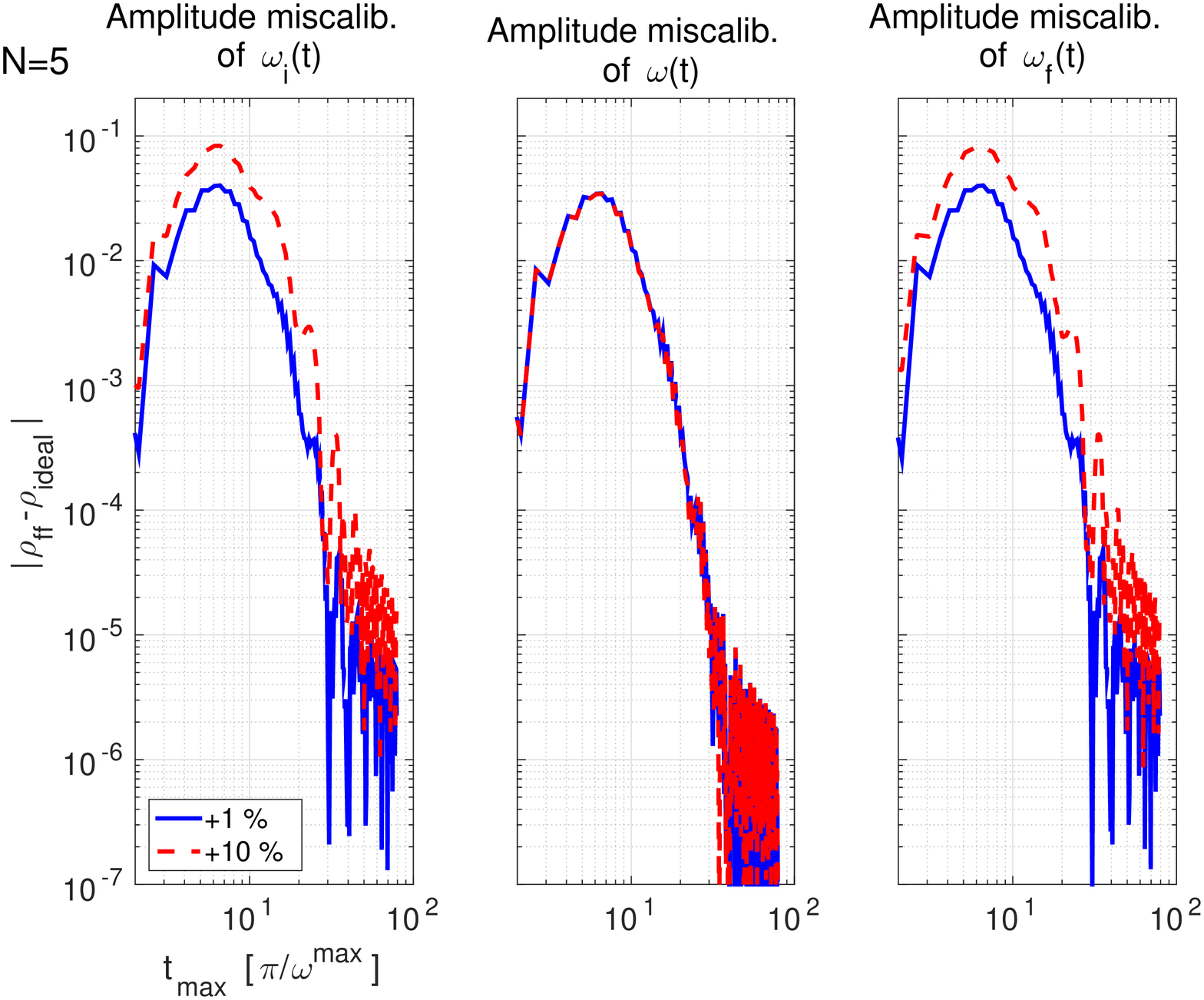}
\end{center}
\caption{Absolute value of the difference between $\rho_{ff}$ and $\rho_{ideal}$ as a function of $t_{max}$ for two different variation (+ 1\% and +10\%) on the amplitude of $\omega_{i}(t)$, $\omega(t)$ and $\omega_{f}(t)$ for $N$=5.}\label{N5amp}
\end{figure}

\begin{figure}[h]
\begin{center}
\includegraphics[width=0.5\textwidth]{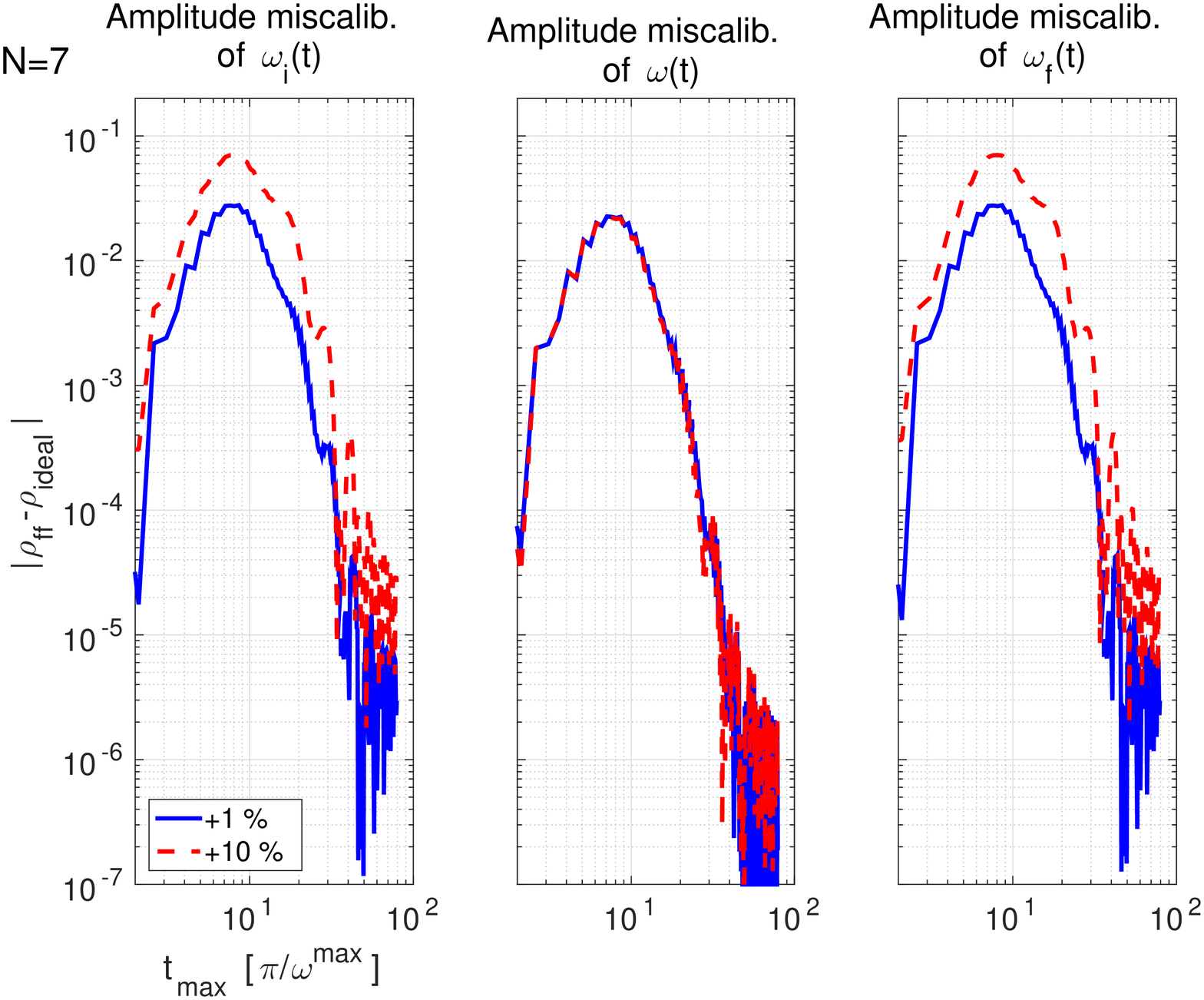}
\end{center}
\caption{Same as Figure \ref{N5amp} with $N$=7.}\label{N7amp}
\end{figure}

\begin{figure}[h]
\begin{center}
\includegraphics[width=0.5\textwidth]{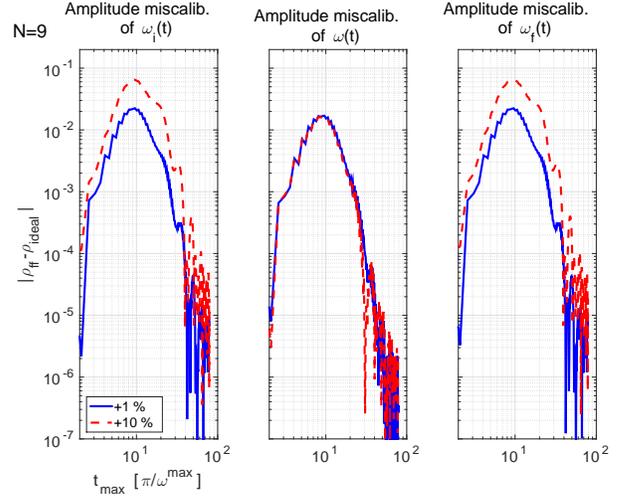}
\end{center}
\caption{Same as Figure \ref{N5amp} with $N$=9.}\label{N9amp}
\end{figure}

In the following, we analyse how a miscalibrated control affects the CTAP, considering deviations in the amplitude and in the peak time of the Gaussian pulses with respect the ideal case.
The analysis methodology is the following: after selecting one of the control signals ($\omega_{i}(t)$ or $\omega(t)$ or $\omega_{f}(t)$) the corresponding miscalibrated pulse is obtained by adding a percentage of the ideal amplitude (peak time) to the selected control signal. Finally, equation (\ref{me}) with the miscalibrated pulse is solved.
In Figures \ref{N3amp}, \ref{N5amp}, \ref{N7amp} and \ref{N9amp} we report the resulting absolute value of the difference between the probability $\rho_{ff}$ for two deviations (+1 \% and +10\%) with respect the amplitude of the ideal Gaussian pulses and the ideal one $\rho_{ideal}$ as a function of $t_{max}$ for $N$=3,5,7 and 9, respectively.
Only one of the signals features an unideal control.

\begin{figure}[h]
\begin{center}
\includegraphics[width=0.5\textwidth]{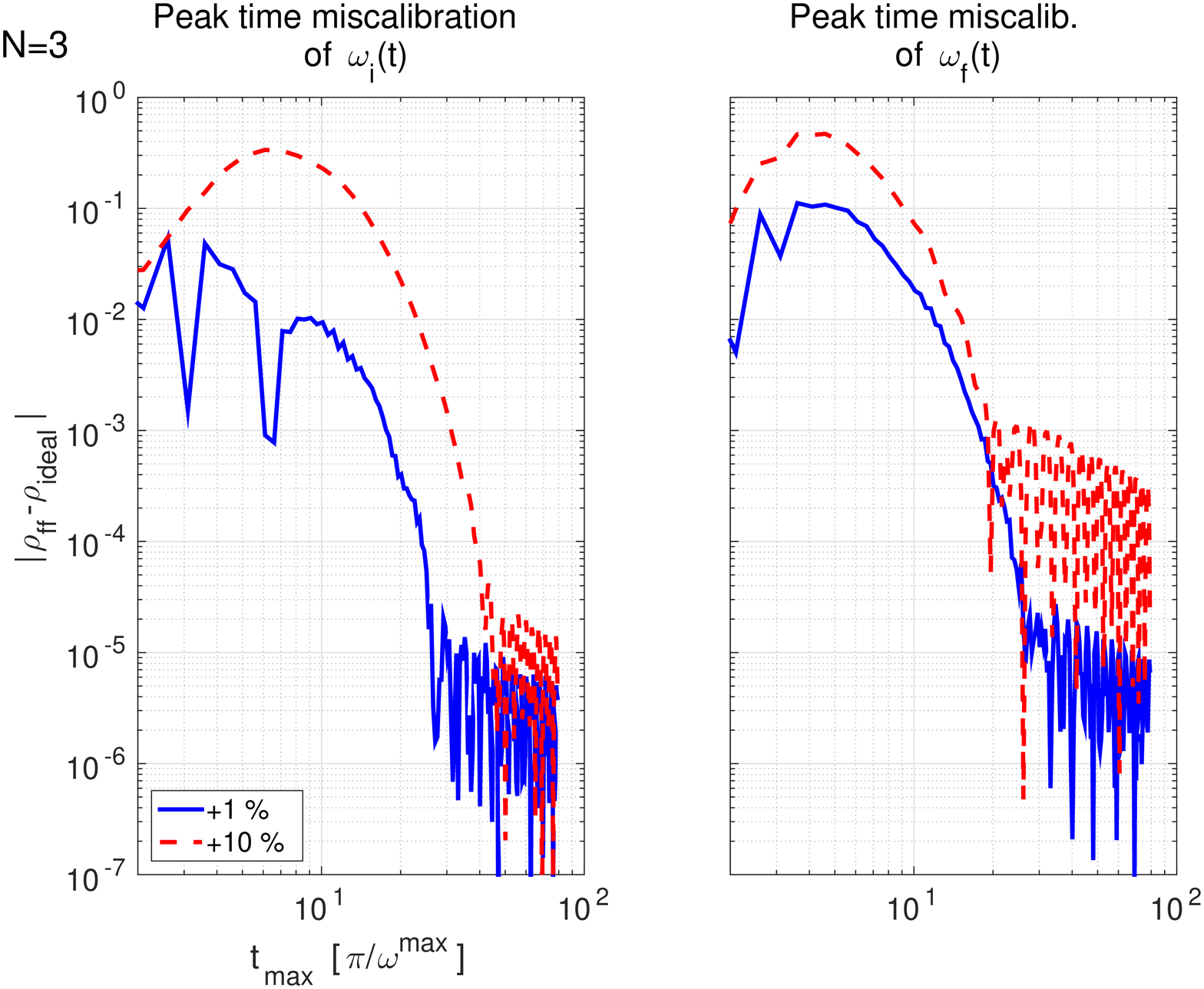}
\end{center}
\caption{Absolute value of the difference between $\rho_{ff}$ and $\rho_{ideal}$ as a function of $t_{max}$ for two different variation (+1\% and +10\%) on the peak time of $\omega_{i}(t)$ and $\omega_{f}(t)$ for $N$=3.}\label{N3time}
\end{figure}

\begin{figure}[h]
\begin{center}
\includegraphics[width=0.5\textwidth]{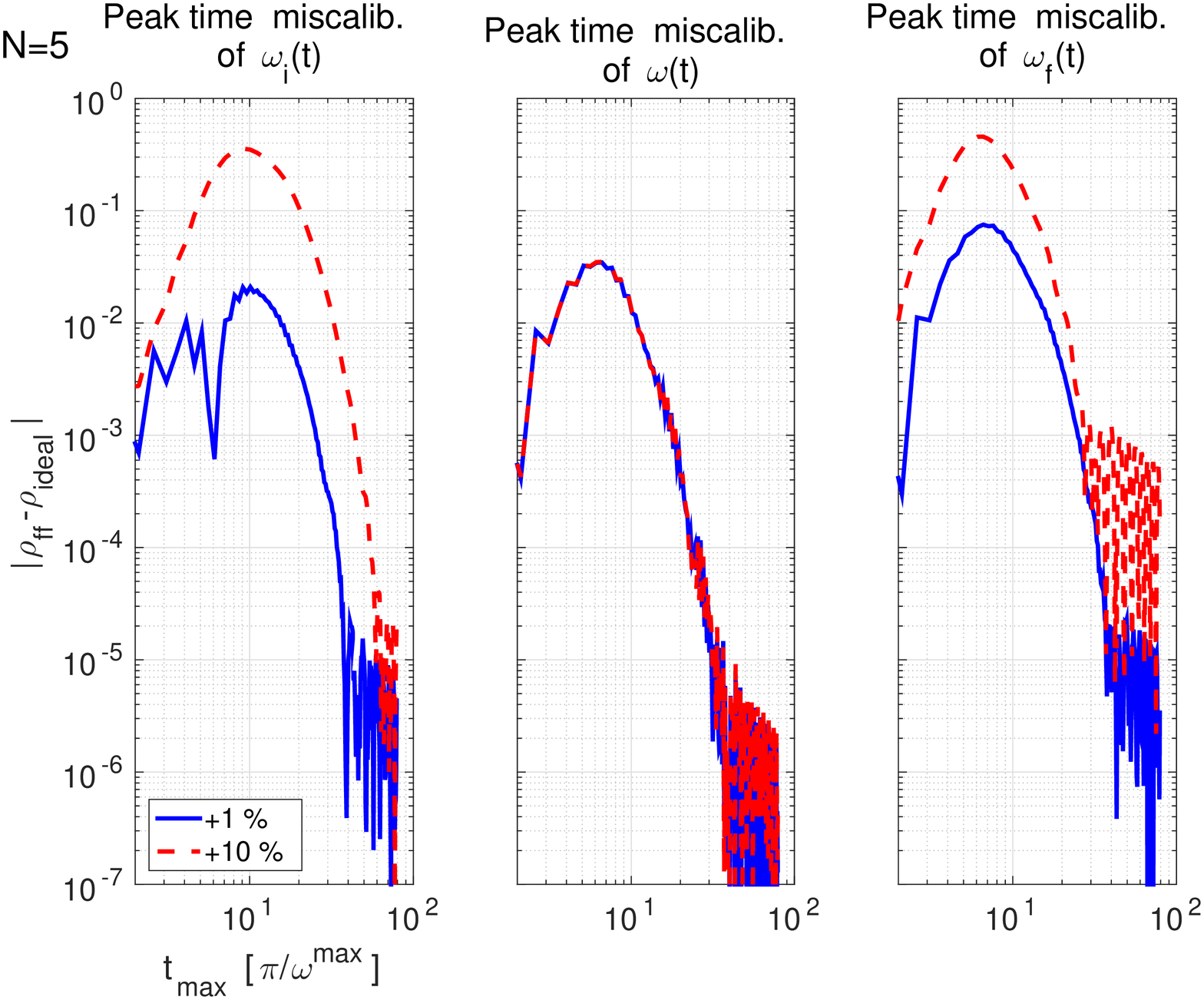}
\end{center}
\caption{Absolute value of the difference between $\rho_{ff}$ and $\rho_{ideal}$ as a function of $t_{max}$ for two different variation (+1\% and +10\%) on the peak time of $\omega_{i}(t)$, $\omega(t)$ and $\omega_{f}(t)$ for $N$=5.}\label{N5time}
\end{figure}

\begin{figure}[h]
\begin{center}
\includegraphics[width=0.5\textwidth]{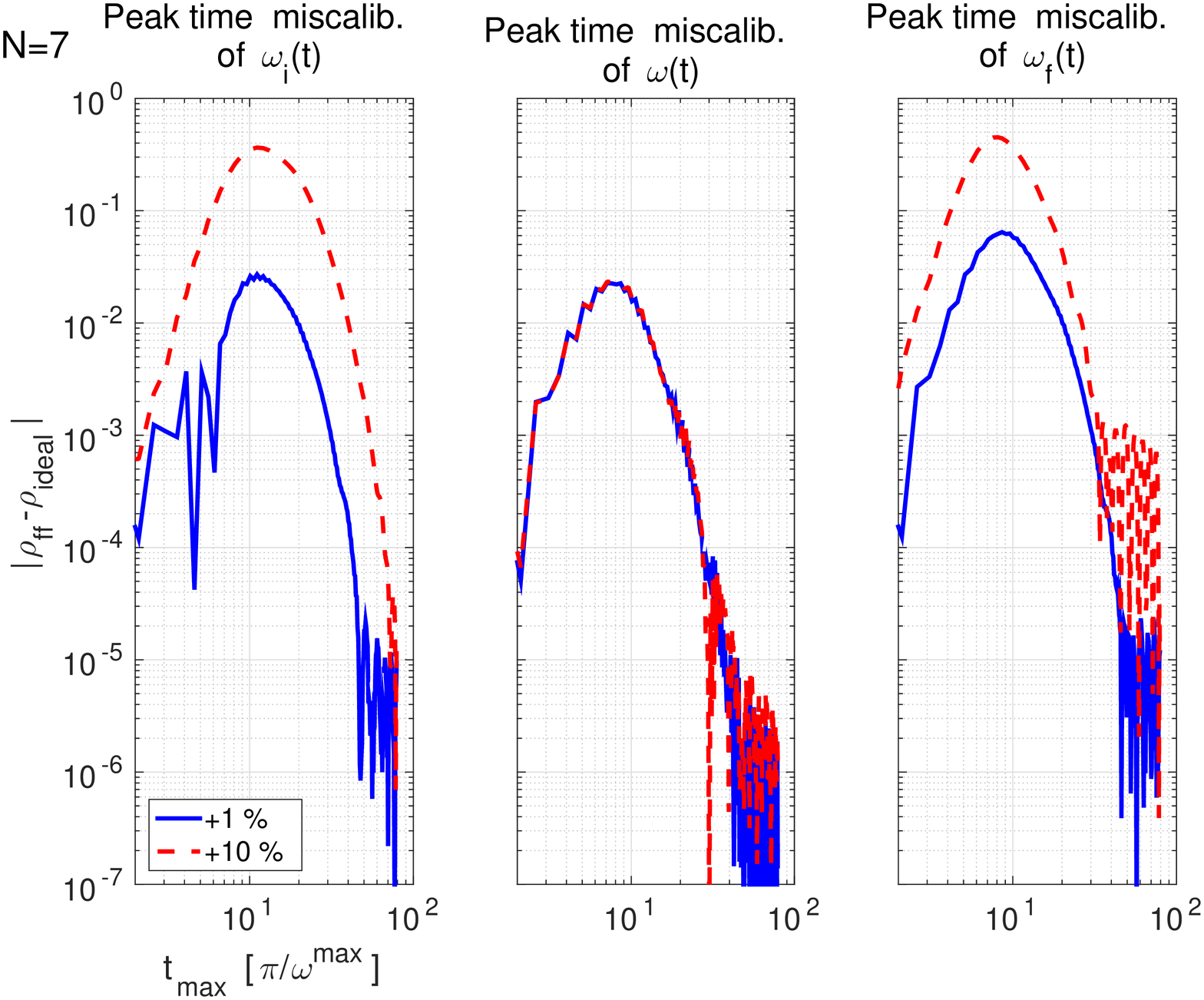}
\end{center}
\caption{Same as Figure \ref{N5time} with $N$=7.}\label{N7time}
\end{figure}

\begin{figure}[h]
\begin{center}
\includegraphics[width=0.5\textwidth]{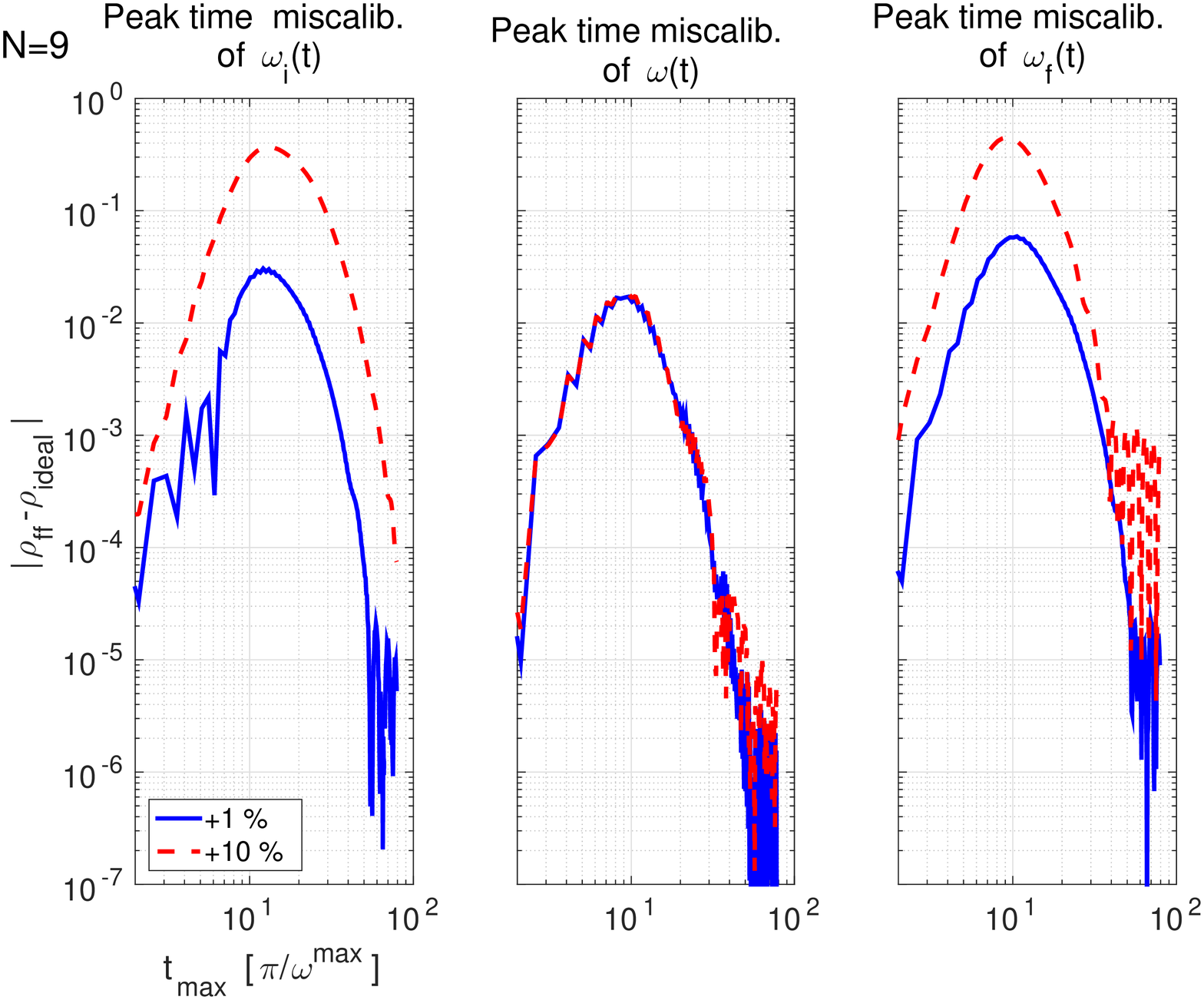}
\end{center}
\caption{Same as Figure \ref{N5time} with $N$=9.}\label{N9time}
\end{figure}

In Figures \ref{N3time}, \ref{N5time}, \ref{N7time} and \ref{N9time}, the quantity $|\rho_{ff}$-$\rho_{ideal}|$ is plotted as a function of $t_{max}$ when one of the control signals features a peak time deviating +1\% and +10\% from the ideal one for $N$ ranging from 3 to 9.

Such results highlight that the CTAP is more sensitive to peak time variations of the Gaussian pulses whereas it better tolerates a miscalibration on the amplitude of the control signals.

\section{Comparison between CTAP and successive SWAPs}
\label{App:CTAPvsSWAP}
\begin{figure}[h]

\begin{center}
\includegraphics[width=0.4\textwidth, clip]{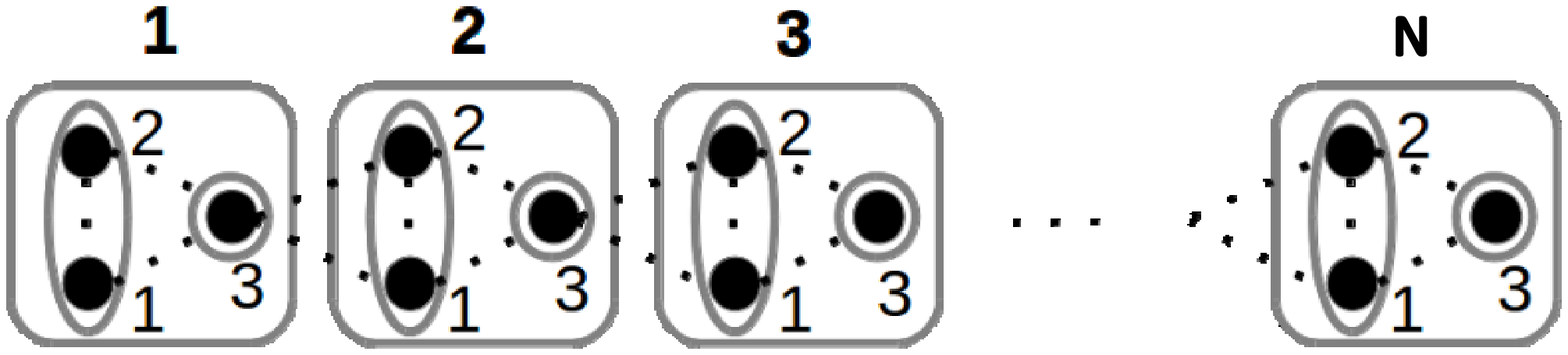}
\end{center}
a)
\begin{center}
\includegraphics[width=0.52\textwidth, clip]{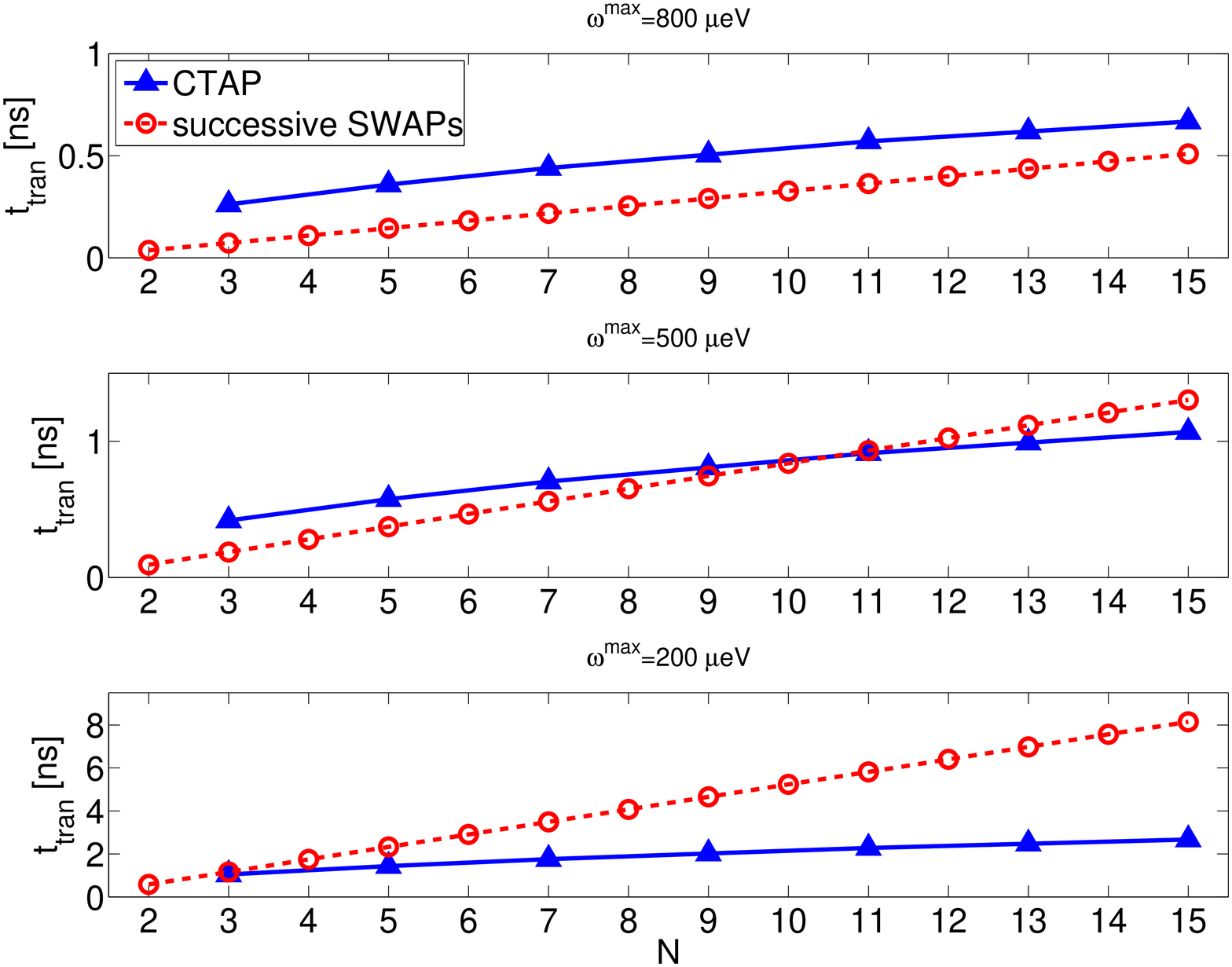}
\end{center}
b)
\caption{a) Scheme of the qubit chain where performing successive SWAP operations. Notat that every qubit must be filled with 3 electrons. b) Comparison between the transfer time for CTAP and successive SWAPs as a function of the chain length $N$ for three different values of the tunnelling rate $\omega^{max}$.}\label{Fig:CTAPvsSWAPchain}
\end{figure}

In the following a comparison between the CTAP and an alternative method to transfer the quantum state of a qubit to a remote one is presented.
The alternative method considered here consists to connect two remote qubits with a chain of qubits and to apply successive SWAP operations (the SWAP exchanges quantum states of two adjacent qubits) to move the quantum state of the head qubit towards the tail qubit. The sequence of effective exchange interactions to perform the SWAP operation between two exchange-only spin-qubits is presented in \cite{RottaQIC2014} with a total sequence time $t_{SWAP}$=11.254 [$h$/$J^{max}$].
In the case of successive SWAPs, the chain of $N$ double QDs is arranged as highlighted in Figure \ref{Fig:CTAPvsSWAPchain}a with a chain transfer time
\begin{equation}
t_{tran}=(N-1) t_{SWAP} \ h\frac{\Delta E_{ST}}{(\omega^{max})^2}
\end{equation}
with $\omega^{max}$ the maximum tunnelling rate between double QDs and $\Delta E_{ST}$ the singlet-triplet energy splitting in the double occupied QDs.
The comparison between the transfer time of CTAP $t_{max}$ and the corresponding one for successive SWAPs $t_{tran}$ is shown in Figure \ref{Fig:CTAPvsSWAPchain}b as a function of number $N$ of double QDs for three different values of the maximum tunnelling rate and with a fixed $\Delta E_{ST}$= 500 $\mu eV$.
The transfer times increase as $N$ grows for both the approaches, with a linear behaviour for the successive SWAPs and a sub-linear one for the CTAP. Given $\Delta E_{ST}$, the tunnelling rate plays a fundamental role in determining which is the fastest approach to be preferred. In fact, successive SWAPs have to be preferred for high tunnelling rates whereas the CTAP is faster for lower rates irrespectively on $N$. When the tunnelling rates are closer to $\Delta E_{ST}$=500 $\mu$eV successive SWAPs are faster for shorter chains whereas CTAP must be preferred for farther interconnections.
Note that an intrinsic disadvantage of the approach with successive SWAPs is the mandatory initial filling of the QDs with electrons which is not required in the case of CTAP (see Figure \ref{Fig1}).

\end{document}